\documentclass[useAMS,usenatbib]{mn2e}
\pdfoutput=1
\usepackage{graphics}
\usepackage{rotating}
\usepackage{ifpdf}
\usepackage{graphicx}

\def\ll{\lambda\lambda}
\def\HeI{\hbox{He~\textsc{I}}}
\def\HeII{\hbox{He~\textsc{II}}}
\def\SiII{\hbox{Si~\textsc{II}}}
\def\SiIII{\hbox{Si~\textsc{III}}}
\def\SiIV{\hbox{Si~\textsc{IV}}}
\def\NIII{\hbox{N~\textsc{III}}}
\def\NIV{\hbox{N~\textsc{IV}}}

\newcommand{\lsim}{\ \raise
-2.truept\hbox{\rlap{\hbox{$\sim$}}\raise5.truept\hbox{$<$}\ }}
\newcommand{\gsim}{\ \raise
-2.truept\hbox{\rlap{\hbox{$\sim$}}\raise5.truept\hbox{$>$}\ }}

\title[The Massive Stellar Population in LH~95]{The Massive Stellar Population in the Young Association LH~95 in the LMC}

\author[N. Da Rio et al.]
  {N.~Da Rio,$^1$\thanks{ndario@rssd.esa.int}
  D.~A.~Gouliermis,$^{2,3}$ B. Rochau,$^2$ A.~Pasquali,$^4$
  J.~Setiawan,$^2$  \newauthor and G.~De~Marchi$^1$
  \\
  $^1$European Space Agency, Keplerlaan 1, 2200 AG Noordwijk, The Netherlands\\
  $^2$Max Planck Institut f\"{u}r Astronomie,  K\"{o}nigstuhl 17, D-69117 Heidelberg, Germany \\
  $^3$Zentrum f\"ur Astronomie der Universit\"at Heidelberg, Institut f\"ur Theoretische Astrophysik, \\
  ~Albert-Ueberle-Str.~2, 69120 Heidelberg, Germany\\
  $^4$Zentrum f\"{u}r Astronomie der Universit\"{a}t Heidelberg, Astronomisches Rechen-Institut, \\
 ~M\"{o}nchhofstr. 12-14, 69120 Heidelberg, Germany}

\begin{document}

\date{MNRAS accepted}

\pagerange{\pageref{firstpage}--\pageref{lastpage}} \pubyear{2012}

\maketitle

\label{firstpage}


\begin{abstract}
We present a spectroscopic study of the most massive stars in the young (4~Myr old) stellar cluster LH~95 in the Large Magellanic Cloud. This analysis allows us to complete the census of the stellar population of the system, previously investigated by us down to 0.4 solar masses with deep HST Advanced Camera for Surveys photometry. We perform spectral classification of the five stars in our sample, based on high resolution optical spectroscopy obtained with 2.2m~MPG/ESO FEROS. We use complementary ground-based photometry, previously performed by us, to place these stars in the Hertzsprung-Russel diagram. We derive their masses and ages by interpolation from evolutionary models. The average ages and age spread of the most massive stars are found to be in general comparable with those previously derived for the cluster from its low mass PMS stars. We use the masses of the 5
sample stars to extend to the high-mass end the stellar initial mass function of LH~95 previously established by us. We find that the initial mass function follows a Salpeter relation down to the intermediate-mass regime at 2~M$_\odot$. The second most massive star in LH~95 shows broad Balmer line emission and infrared excess, which are compatible with a classical Be star. The existence of such a star in the system adds a constrain to the age of the cluster, which is well covered by our age and age spread determinations. The most massive star, a 60-70~M$_\odot$ O2 giant is found to be younger ($<1$~ Myr) than the rest of the population. Its mass in relation to the total mass of the system does not follow the empirical relation of the maximum stellar mass versus the hosting cluster mass, making LH~95 an exception to the average trend.
\end{abstract}




\section{Introduction}
It is well established \citep{ladalada2003} that the vast majority of stars are not formed in isolation, but as part of stellar associations and clusters. These comprise hundreds or even thousands of stars along the entire mass spectrum, from the most massive OB-type down to low-mass late-type stars. The study of young (few Myr old) clusters, therefore, provides a fundamental tool to understand how the star formation process takes place. Both high- and low-mass stars are, from different points of view, of complementary importance in such clusters. On one hand, the short-lived massive stars are the direct evidence of the youth of these systems, and their energy output through winds and UV radiation is responsible for the gas removal, a process that strongly affects the survival of the stellar system. On the other hand, the low-mass stars constitute the majority of the newly formed stellar population, and their slow evolution during the pre-main sequence (PMS) phase traces the timescale of star formation.

The study of star-forming regions in the Milky Way is affected by the high extinction produced by the dust in the Galactic disk, limiting the depth of optical observations. This problem does not affect significantly studies in the Magellanic Clouds (MCs), our nearby metal-poor dwarf galaxies. Both the Large and Small Magellanic Clouds (LMC, SMC) are characterized by an exceptional sample of star-forming regions, the {\sc Hii} regions \citep[][]{henize56, davies76}, where hydrogen is already being ionized by the UV-winds of massive stars. A plethora of young stellar associations and clusters are embedded in these regions  \citep[e.g.,][]{bica95, bica99}, giving evidence  of current clustered star formation.

LH~95 \citep{lucke70} is a young stellar cluster in the LMC related to the {\sc Hii}  region LHA 120-N 64 \citep{henize56}, located to the north-east of
the super-bubble LMC~4. The association has a projected size of $\sim10\times20$~pc, and is not centrally concentrated but presents at least 3 main subclusters.
Its stellar population has been studied by \citet{gouliermis02} using ground-based $BVR$ and H$\alpha$ photometry; they determined the initial mass function (IMF) from $30$~M$_\odot$ to $2$~M$_\odot$.
More recently, very deep Hubble Space Telescope \emph{Advanced Camera for Surveys} (HST/ACS) observations in $F555W$ and $F814W$ bands \citep{gouliermis2007} led to the discovery of a population of $\sim2500$ PMS stars in the region. In \citet[][hereafter, Paper~I]{DaRio2009}, using these data, we derived the first extragalactic IMF in the subsolar regime, complete down to $\sim 0.4 M_\odot$ while members have been detected with masses as low as $0.2 M_\odot$. Later, in \citet[][hereafter Paper~II]{DaRio2010}, we investigated the age and age spread of the intermediate- and low-mass stars in LH~95 through a statistical technique that accounts for physical and observational biases in the observed colors and magnitudes. We found an average age of the system of the order of 4 Myr, and an age spread well represented by a gaussian distribution with a standard deviation of $1.2$ Myr.

Despite the large luminosity (and mass) range spanned by our ACS observations, our stellar sample was incomplete to the high-mass end, since five of the brightest stars were saturated in the HST images.
Moreover, optical photometry of high-mass stars (with $M\gsim8$~M$_\odot$) alone is not sufficient for the accurate derivation of their stellar parameters. This is because these stars, due to their high effective temperatures $T_{\rm eff}$, emit a large fraction of their flux in the UV wavelength range. As a consequence, the optical bands are in the Rayleigh regime and the observed colors tend to be independent of $T_{\rm eff}$.  This well-known problem was made evident also in Paper~II, where we demonstrated that dating massive stars based only on their photometry is highly uncertain.

Stellar parameters of massive stars can be determined accurately through spectroscopy. Therefore, in order to study the brightest stellar members of LH~95,
and to construct the H-R diagram of its massive stellar population, we perform here a spectroscopic follow-up to our previous studies. We acquired optical spectra of
the five brightest stars in the region and we complemented these data with existing $BVR$ photometry of LH~95, obtained from the ground \citep{gouliermis02}. The  aim of this investigation is twofold: 1) the completion of the IMF derived in Paper~I toward the highest masses, and 2) the analysis of the ages of the high-mass members (and possible variations), in relation to the similar results we presented in Paper~II for the low-mass population.

This paper is organized as follows. The spectroscopic observations and their data reduction are described in Sections \ref{section:the_observations} and \ref{section:data_reduction}. In Section \ref{section:spectral_classification} we present the classification of the sources. In Section \ref{section:HRD} we derive the Hertzsprung-Russel diagram for the high-mass end of the LH~95 population, and assign ages and masses to the sources. A discussion on these results is presented in Section \ref{section:discussion}, and in Section \ref{section:IMF} we discuss the high-mass end of the IMF. We then focus in Section \ref{section:Be_star} on the brightest source in the sample, where we present evidence that classifies it as a classical Be star. Finally, in Section \ref{section:conclusions} we summarize the main findings of this paper.


\section{The Observations}
\label{section:the_observations}
\begin{figure*}
\includegraphics[width=2\columnwidth]{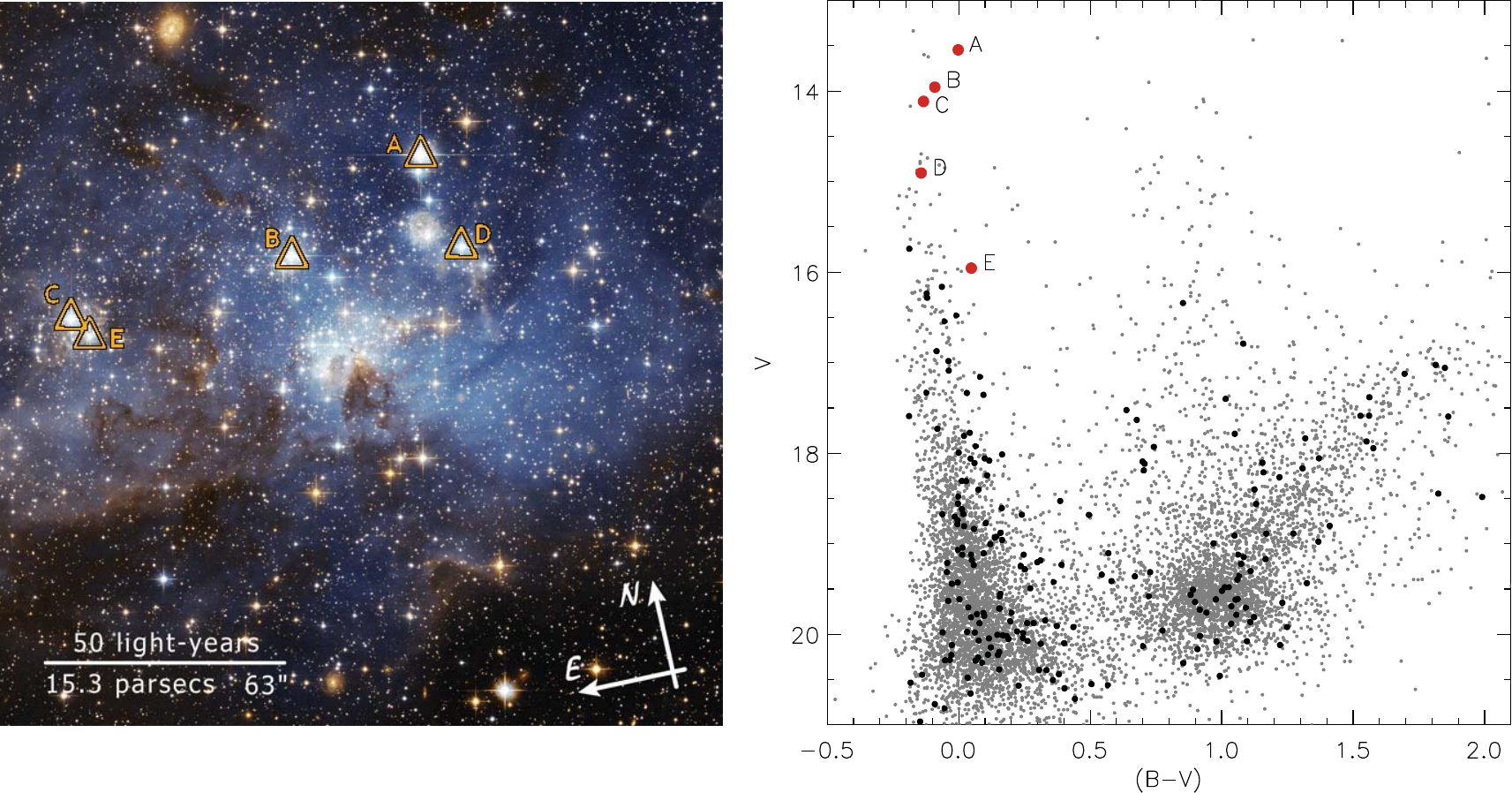}
\caption{\emph{Left panel:} The five targets selected for our spectroscopic follow-up, onto the color-composite image of LH~95 from ACS observations \emph{(Image credit: NASA, ESA and D. A. Gouliermis -- MPIA)}. \emph{Right panel:} CMD from the \citet{gouliermis02} photometry. The black dots indicate sources located within the ACS FOV. The five targets selected for spectroscopy are marked with thick red dots.\label{fig:FOV}}
\end{figure*}

This spectroscopic study focuses on the brightest stellar sources in LH~95, which were saturated in our HST/ACS images. For the photometry of these sources we use the measurements by \citet{gouliermis02}, based on $BVI$ ground-based observations. While this photometric study achieved significantly shallower detection limits and poorer spatial resolution than our ACS imaging, it provides accurate $BVI$ photometry for all the bright sources within the star-forming region. Based on this study we excluded from our analysis the bright stars with significantly red colors, as probably being field RGB stars. This is because at the age of LH~95 ($\sim 4$~Myr), the turn-off from the main sequence occurs at a large mass ($\sim 40$~M$_\odot$), and the RGB stars above this mass are not expected to have temperatures lower than 6000~K \citep{girardi02,schaerer+1993}, thus should not show very red optical colors. The remaining five bright main-sequence stars are the targets of this spectroscopic follow-up. We annotate these sources with the letters A, B, C, D, and E in the color-composite $VI$ image of our HST/ACS observations, shown in Figure~\ref{fig:FOV} (left panel). The $B-V$, $V$ color-magnitude diagram (CMD) from the photometric study of \citet{gouliermis02} is shown also in Figure \ref{fig:FOV} (right panel). In this plot  stars identified by \citet{gouliermis02} in a significantly larger field-of-view (FOV) than that of ACS are plotted with grey symbols. The stars observed within our ACS FOV \citep{gouliermis2007} are highlighted with large black symbols, and the targets of this spectroscopic study with red symbols. They are also annotated in this CMD with their alphabetic IDs.

The observations have been carried out with the Fibre-fed, Extended Range \`{E}chelle Spectrograph (FEROS), mounted on the 2.2 m MPG/ESO telescope located at the La Silla Observatory in Chile. This instrument has a resolution of $R=48,000$ and a wavelength coverage of $3600$~-~$9200$ \AA\ \citep{kaufer98}. The spectrograph is fed by two fibres with an aperture of 2.0 arcsec, providing simultaneous spectra of a stellar target, plus either the sky or one of two calibration lamps (wavelength calibration and flat-field). Since for our purposes the wavelength calibration is not critical, we used the ``OBJSKY'' instrument setup, acquiring a position in the sky in parallel to the targets separated by 2.9 arcmin.
The necessary calibration acquisitions have been carried out separately for each observing night.
Observations were carried out in two epochs: November 2008 and January 2009.
The summary of the observations is given in Table \ref{t:obs}. Multiple spectra have been taken for the same targets to enable cosmic ray detection and removal. The overall exposure times have been higher for fainter sources in order to reach a sufficient signal-to-noise ratio (SNR). Specifically, our final spectra reach an average SNR in the optical range spanning from $\sim100$ for star A, to $\sim30$ for star E.

\begin{table*}
\centering
\begin{minipage}{140mm}
\caption{Summary of the observations}
\label{t:obs}
\begin{tabular}{rrrrrrrrr}
\hline
ID & ID2 (G02)\footnote{catalog numbers of Gouliermis et al. (2002)} & alternative name\footnote{Sanduleak (1970) identification numbers} & RA & Dec & V & exposure time & airmass & epoch \\
 & & & (J2000.0) & (J2000.0) & [mag] & [s] & & \\
\hline
A & 9136 & SK -66 170 & 05:36:58.945 & -66:21:16.130 & 13.17 & 3600 &     1.31 & Nov 2008\\
 &  &  &  &  & & 1000   &    1.35 & Nov 2008\\
 &  &  &  &  & & 1500   &    1.39 & Nov 2008\\
 &  &  &  &  & & 1200   &    1.26 & Jan 2009\\
 &  &  &  &  & & 1200   &   1.25 & Jan 2009\\
B       & 9141 & SK -66 172 & 05:37:05.553 & -66:21:34.950 & 13.58 & 1200 &     1.30 & Nov 2008\\
 &  &  &  &  & &  900 &     1.32 & Nov 2008\\
C       & 9143 & SKÂ -66 174 & 05:37:15.723 & -66:21:38.355 & 13.74 & 900 &     1.27 & Nov 2008\\
 &  &  &  &  & & 900 &     1.28 & Nov 2008\\
D       & 18 &  &  05:36:58.007 & -66:21:42.613 & 14.53 & 1000 &     1.25 & Nov 2008\\
 &  &  &  &  & & 1000 &     1.26 & Nov 2008\\
 &  &  &  &  & & 1800 &     1.26 & Jan 2009\\
 &  &  &  &  & & 1800 &     1.28 & Jan 2009\\
E       & 85 &  & 05:37:15.129 & -66:21:44.304 & 15.58 & 1400 &     1.28 & Nov 2008\\
 &  &  &  &  & & 1400 &     1.31 & Nov 2008\\
 &  &  &  &  & & 1400 &     1.34 & Nov 2008\\
 &  &  &  &  & & 1800 &     1.31 & Jan 2009\\
 &  &  &  &  & & 1800 &     1.35 & Jan 2009\\
\hline
\end{tabular}
\end{minipage}
\end{table*}

\section{Data Reduction}
\label{section:data_reduction}

\begin{figure*}
\includegraphics[width=1.8\columnwidth]{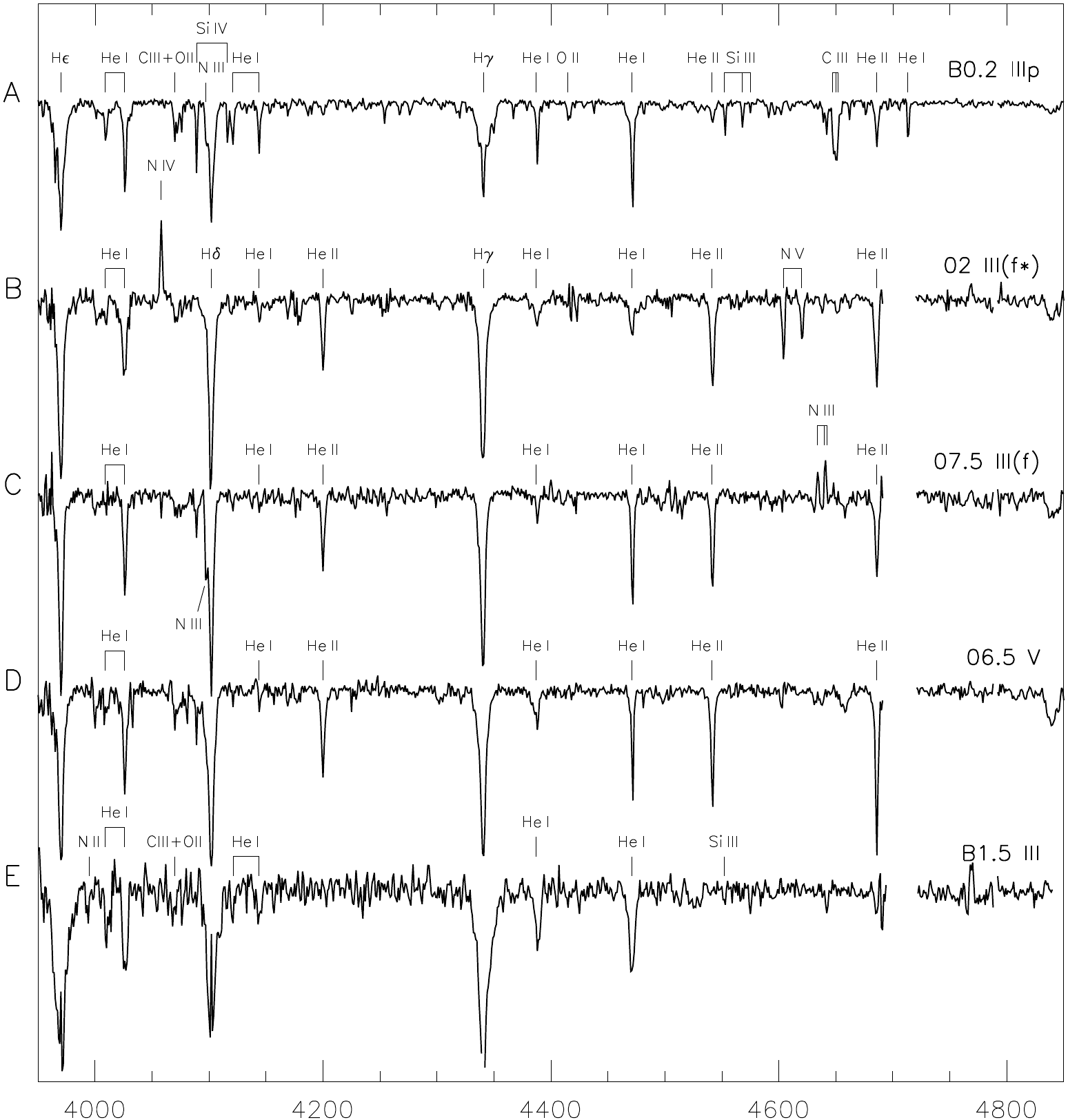}
\caption{Reduced spectra for the five targets, covering the violet-blue range of the spectrum, relevant for spectral classification of early-type stars. The main stellar lines we have identified in the spectra are indicated as well. The features identified in the spectra are: He~\textsc{I} $\lambda\lambda4009, 4026, 4121, 4144, 4387, 4471, 4713$; He~\textsc{II} $\lambda\lambda4200, 4541, 4686$;  C~\textsc{III}/O~\textsc{II} $\lambda4070$; Si~\textsc{IV} $\lambda\lambda4089-4116$;  Si~\textsc{III} $\lambda\lambda4552-4568-4575$; N~\textsc{II} $\lambda3995$; N~\textsc{III} $\lambda\lambda4097, 4634-4640-4642$; N~\textsc{IV} $\lambda4058$;  N~\textsc{V} $\lambda\lambda4604-4620$;  O~\textsc{II} $\lambda4415$; C~\textsc{III} $\lambda\lambda4647-4650-4652$.
\label{fig:spectra}}
\end{figure*}

The basic data reduction of the spectra was performed by using the automatic online pipeline. This pipeline extracts the 39 cross-dispersed spectral orders and combines them together in a single, wavelength calibrated, spectrum. We manually subtracted the relevant sky emission from each spectrum; then we normalized the continuum of each spectrum to unity by fitting a high-order polynomial to the continuum smoothed with a median filter 12\AA\ in size.
Each derived continuum has been visually inspected together with its original spectrum to ensure a correct estimation. The continuum normalization also allows us to correct for shape variations in the spectra of the same star introduced by changes in airmass and atmospheric extinction, as well as flux factors due to different exposure times.
Cosmic rays are detected as $>3\sigma$ peaks above the continuum on each spectrum and absent in the other spectra of the same star. The average local noise, $\sigma$, was computed in a 2~\AA\ neighborhood of each wavelength point. The cosmic rays were then masked, and the spectra were combined in one final spectrum with the use of a weighted average method, accounting for the individual exposure times.
In order to facilitate the spectral classification, the original spectra were also degraded to a resolution of $1\AA$.

The blue wavelength range of the final spectra are presented in Figure \ref{fig:spectra}. In this plot the spectra are blue shifted by an amount $\delta\lambda / \lambda=10^{-3}$, to correct for the radial velocity of the system, $v_r\simeq 302.8$~km~$s^{-1}$, determined from the lines in the spectra. For the four faintest stars, we have manually removed part of the spectra at $4695\AA \leq \lambda \leq 4720\AA$, where an artificial feature in emission dominates the continuum of the 16th spectral order. By visual inspection on the individual spectral orders, we verified that the $\HeII$ $\lambda4686$ line is not affected by this artifact, being located at the red end of the 15th order.

\section{Spectral classification}
\label{section:spectral_classification}

We assign spectral types to our targets by comparing their spectra to stellar atlases available in the literature. In particular we consider the atlases of \citet{walborn1990} and \citet{sota2011}, which contain a large selection of early type stars and provides a fine coverage of the spectral types and luminosity classes of OB stars. We also use the atlases of OB type stars of \citet{walborn95} and \citet{walborn2002}. Based on our classification, we assume an uncertainty in the assigned spectral types of  half a spectral subtype for stars A to D and one subtype for star E.

\subsubsection*{Star A}
This is the brightest stellar source in the region ($V\sim13.5$~mag; \citealt{gouliermis02}). Due to the long total exposure time allocated to this source (exceeding 2 hours) the derived spectrum reaches a very high SNR ($>100$).
The line ratio between \HeI~$\lambda4471$ and  \HeII~$\lambda 4541$, together with the ratio \SiIII\ $\lambda4552$/\SiIV\ $\lambda4089$ indicate a B0.2 type star.
As far as its luminosity class is concerned the spectrum of this star shows some inconsistencies. In particular, while the significant strength of the \SiIV\ lines suggests a very luminous star of class I or II (also considering the low metallicity of the LMC), the presence of the \HeII~$\lambda4686$ line is consistent only with a luminosity class IV or V. Therefore, we assign to this source a giant luminosity class and we classify this star as a B0.2~IIIp. As we discuss later (see Section \ref{section:Be_star}) this star shows prominent H$\alpha$ and H$\beta$ emission, behaving as a classical Be star.

\subsubsection*{Star B}
This star was already classified by \citet{walborn95} as an O3~III(f*), and then reclassified in \citet{walborn2002} as O2~III(f*), with evidence in support of a binary system with a later OB-type star. Our spectrum confirms this classification, mainly based on the prominent \NIV~ $\lambda4058$ emission, the NV\ $\ll4604-4620$ absorption, and the absence of \NIII~$\lambda4640$ emission. The significant strength of the \NIV~ $\lambda4058$ line indicates a luminous star ($\leq$ III), but the \HeII\ $\lambda4686$ in absorption excludes a supergiant. According to \citet{walborn2002}, the spectrum is a composite including an unresolved later-type star, as revealed by the detection of the \HeI\ $\lambda4471$. In our spectrum (see Figure \ref{fig:spectra}) we also clearly detect \HeI\ $\lambda4387$, which did not appear in \citet{walborn95} spectrum. This further supports the claim that this source is a composite system.
For an additional discussion on the composite nature of this source, we refer to Section \ref{section:binarity-starB}.

\subsubsection*{Star C}
According to the ratio between \HeII~$\lambda4541$ and \HeI~$\lambda4471$ (which are equally strong for O7 stars) this source is an O7.5 type. The presence of \NIII~$\ll4634-4640-4642$ emission is typical of a luminous star of this spectral type, and their strength suggests at least a II(f) luminosity class. The strong  \NIII~$\lambda4097$ absorption is also supportive of a bright giant. We also notice a modest emission in the C~\textsc{III} $\lambda\lambda4647-4650-4652$, which is atypical for this spectral type. The strength of the \HeII~$\lambda4686$ absorption in contrast with the \NIII emission, indicating a luminosity not higher than class III. We argue that the unusual strength of the N emission lines might be due to nitrogen enrichment, or, alternatively, could be indicative of a composite nature for this source. As a compromise, we classify this star as O7.5~III(f).

\subsubsection*{Star D}
The ratio between \HeII~$\lambda4541$ and \HeI~$\lambda4471$ indicates a type of O6.5 for this star. In addition, the presence of strong \HeII~$\lambda4686$ absorption, typical of main sequence O-type stars, leads to the classification of this star as an O6.5~V.

\subsubsection*{Star E}
Given the strength of the \HeI\  and the absence of \HeII\ lines, this source is an early B-type star. The strength of \SiIV\ and \SiIII, together with the absence of \SiII~ $\lambda\lambda 4128-4130$ indicate a B1.5 star. The additional presence of the N~\textsc{II} $\ll 3995$ suggests a luminous star, but not a supergiant given the absence of several indicators of such high luminosity (e.g., strong O~\textsc{II} lines). Therefore we assign the spectral type B1.5~III.

\section{The H-R diagram}
\label{section:HRD}

In this section we derive the stellar parameters of the observed sources from both our spectroscopic analysis described above, and the available photometric measurements of these sources \citep{gouliermis02}, and we construct the H-R diagram for these stars.
We consider appropriate temperature scales for early-type stars with LMC metallicity ([Fe/H]$\simeq-0.3$; \citealt{luck98}), and we derive the effective temperatures ($T_{\rm eff}$) of the stars from their spectral types.
There is a number of spectral type - $T_{\rm eff}$ conversions available in the literature, covering different ranges in the parameter space. The derivation of such relations usually relies on model stellar atmospheres, in which $T_{\rm eff}$ is an input parameter. The associated spectral types are obtained
from the presence of diagnostic lines, following observational criteria such as those we applied in the previous section.

The stellar
atmosphere modeling in particular for O-type stars is quite challenging \citep[see, e.g.,][]{kudritzkipuls00}, since it requires several factors to be considered for the effects of metals in the atmospheric structure and emergent spectrum, such as a non-local thermodynamic equilibrium (NLTE) treatment, spherical expansion due to stellar winds, and line blanketing \citep{martins+2005}. Such studies show a clear dependence of the temperature scale on the luminosity class, with luminous stars being colder than dwarfs at a given spectral type. In addition, \citet{massey+2005} found evidence for a systematic change in the temperature scale with metallicity. They collected high SNR optical and UV spectra of a sample of O-type stars in the Galaxy and in the MCs, and derived the stellar $T_{\rm eff}$ via modeling. These authors found that SMC O-type stars are systematically hotter than Galactic stars at the same spectral type, while LMC stars follow an intermediate temperature scale between the two, as expected by their metallicity. In agreement with previous works, they also found lower $T_{\rm eff}$ for supergiants (luminosity class I), while no significant differences were measured between giants and dwarfs (III and V).

\begin{table*}
\centering
\begin{minipage}{1.55\columnwidth}
\caption{Derived stellar parameters for our target stars.}
\label{t:param}
\begin{tabular}{crrrrcc}
\hline
ID & $T_{\rm eff}$ & $BC_V$ & $A_V$ & ${\log L/L_\odot}$\footnote{The first error in $\log L/L_{\odot}$ represents the propagation of the uncertainty in $\log T_{\rm eff}$ (and consequently in $BC$), the second error represents the propagation of the uncertainty in $M_V$.} & mass\footnote{The masses and ages of the stars are determined in Section~\ref{s:massage}.} & age \\
&(K) & (mag) & (mag) & & (M$_{\odot}$) & (Myr) \\
\hline
 A&      $29170^{+270}_{-330}$   & $-2.816^{-0.066}_{+0.046}$ & $0.880\pm0.174$ & $5.474^{+0.026\pm0.07}_{-0.018\pm0.07}$ & $33.2\pm2.7$ & $5.43\pm0.41$ \\ \\
 B&      $50000^{+2000}_{-2000}$ & $-4.433^{-0.118}_{+0.118}$ & $0.644\pm0.174$ & $5.862^{+0.047\pm0.07}_{-0.047\pm0.07}$ & $72.1\pm6.9$ & $0.33\pm0.18$ \\ \\
 C&      $35480^{+660}_{-650}$   & $-3.407^{-0.057}_{+0.060}$ & $0.464\pm0.174$ & $5.316^{+0.023\pm0.07}_{-0.024\pm0.07}$ & $31.0\pm2.1$ & $4.55\pm0.18$ \\ \\
 D&      $38110^{+800}_{-610}$   & $-3.620^{-0.058}_{+0.050}$ & $0.433\pm0.174$ & $5.073^{+0.023\pm0.07}_{-0.020\pm0.07}$ & $26.8\pm1.3$ & $2.05\pm1.14$ \\ \\
 E&      $24380^{+1990}_{-4330}$ & $-2.430^{-0.340}_{+0.490}$ & $0.982\pm0.174$ & $4.396^{+0.136\pm0.07}_{-0.196\pm0.07}$ & $13.0\pm1.1$ & $14.7\pm2.5$ \\
\hline
\end{tabular}
\end{minipage}
\end{table*}

Considering these results, for the late O-type stars in our sample we adopted the \citet{massey+2005} LMC scale, which covers the range O3 - O9.5. Temperatures are not explicitly provided for O2 stars, but the sample of \citet{massey+2005} includes two LMC O2~III stars, to which they assign a lower-limit of $T_{\rm eff}=48000$~K, and two O2~V with temperatures between 51000~K and 55000~K. Based on these measurements, we assign a $T_{\rm eff}=50000$~K$~\pm2000$~K to our O2~III star. This value is also compatible with the Massey V+III scale extrapolated to  O2 stars and in agreement with other O2  giant stars in the LMC \citep[e.g.,][]{evans+2010}.
For the two early B-type stars, we adopt the temperature scale of \citet{bessel1998}, which matches very well the Massey scale at B0. We note that for B-type stars, the uncertainties in the spectra modeling become less prominent. This is supported by the fact that all the recent temperature scales for O-type stars \citep[e.g.,][]{vacca+1996,martins+2005,massey+2005} are in agreement with each other at the O-B transition predicted at $T_{\rm eff}\simeq30000$~K, whereas they present significant differences at higher $T_{\rm eff}$. Moreover, in the B-type range we do not expect any significant dependence of the $T_{\rm eff}$ scale on metallicity, as shown by \citet{massey+2005}.

We derive the bolometric luminosities of our stars by first correcting our $V$ magnitudes for extinction, and then applying bolometric corrections according to the stellar temperatures. Extinctions are estimated from the $E(B-V)$ color excess in the photometry of \citet{gouliermis02}. We use the intrinsic $(B-V)_0$ colors published by \citet{martins&pletz2006} for O-type stars, and by \citet{bessel1998} for B-type stars. Since O-type stars are well into the Rayleigh-Jeans regime at optical wavelengths, the intrinsic colors are hardly sensitive on $T_{\rm eff}$. According to \citet{martins&pletz2006}, $(B-V)_0$ decreases from 0.28~mag to only 0.26~mag in the entire O spectral class, with differences between  luminosity classes less than 0.01~mag. Subsequently, we also assume that metallicity variations do not affect the intrinsic $(B-V)$ color. For the two early-B stars in our sample we consider the intrinsic colors from \citet{bessel1998}, assuming $\log g=4.0$. According to this study, at  the $T_{\rm eff}$ of early B-type stars, changes in surface gravity of more than 1~dex cause very small variations of $(B-V)_0$, of the order of 0.01~mag. As a consequence, the uncertainties in the luminosity classes of sources A and E are not expected to bias in any way the assumed intrinsic colors of the two stars, and thus the derived optical extinction.

The measured color excess $E(B-V)$ is then converted into $A_V$ assuming the average LMC reddening law $R_V=3.41\pm0.06$ \citep{gordon+2003}, a value  slightly higher than the typical Galactic, $R_V=3.1$ \citep{cardelli}.
The photometric errors in $B$ and $V$ are negligible, given the high luminosity of the bright members of LH~95. Nevertheless, we add a constant 0.035~mag uncertainty to every star in each band, representing the precision of the photometric calibration of \cite{gouliermis02}. Therefore, the derived $E(B-V)$ has an error of $0.05$~mag.
We convert the extinction corrected magnitudes, $V_0$, into absolute magnitudes, $M_V$, considering a distance modulus $(m-M)_0=18.41$~mag (Paper~I).

Finally, we convert absolute magnitudes into total stellar luminosities ($\log L/L_\odot$), by applying the appropriate $V-$band bolometric corrections ($BC_V$). For our O-type stars we use the relation from \citet{massey+2005}: $$BC=27.99-6.90\cdot\log T_{\rm eff}.$$ Clearly, as for the intrinsic colors, these authors found BCs to be mainly dependent on temperature, and independent on luminosity. Also, they found good agreement with the BCs derived from previous works \citep{vacca+1996}. A comparison of this relation with the bolometric corrections of \citet{martins+2005} yields  differences below 2\%. For the BCs of B-type stars, we consider again the values from \citet{bessel1998}.

Star A, as we describe in detail in Section \ref{section:Be_star}, appears to be a Be star, showing prominent H$\alpha$ and H$\beta$ emission. This excess, however, does not affect our derived stellar parameters for this star. The H$\alpha$ line is at wavelengths longer than the $V$-band bandpass; on the contrary, the H$\beta$ is inside the $B-$band throughput. As we show in Figure \ref{fig:starA_balmer}, however, this line emission appears to be modest, with an equivalent width (E.W.) of $\sim2.8\AA$. This is small compared to the E.W. of the $B-$band filter profile, which is $\sim912\AA$. Furthermore, at the H$\beta$ wavelength ($\lambda=4861\AA$) the filter throughput is about 40\% its peak. As a consequence, the line emission we detect for this star produces a measured flux excess in our $B-$band photometry of the order of 0.1\%, or 0.001~mag. This is therefore absolutely negligible for the derivation of the $A_V$ from the $E(B-V)$ color excess.

\begin{figure*}
\includegraphics[width=1.4\columnwidth]{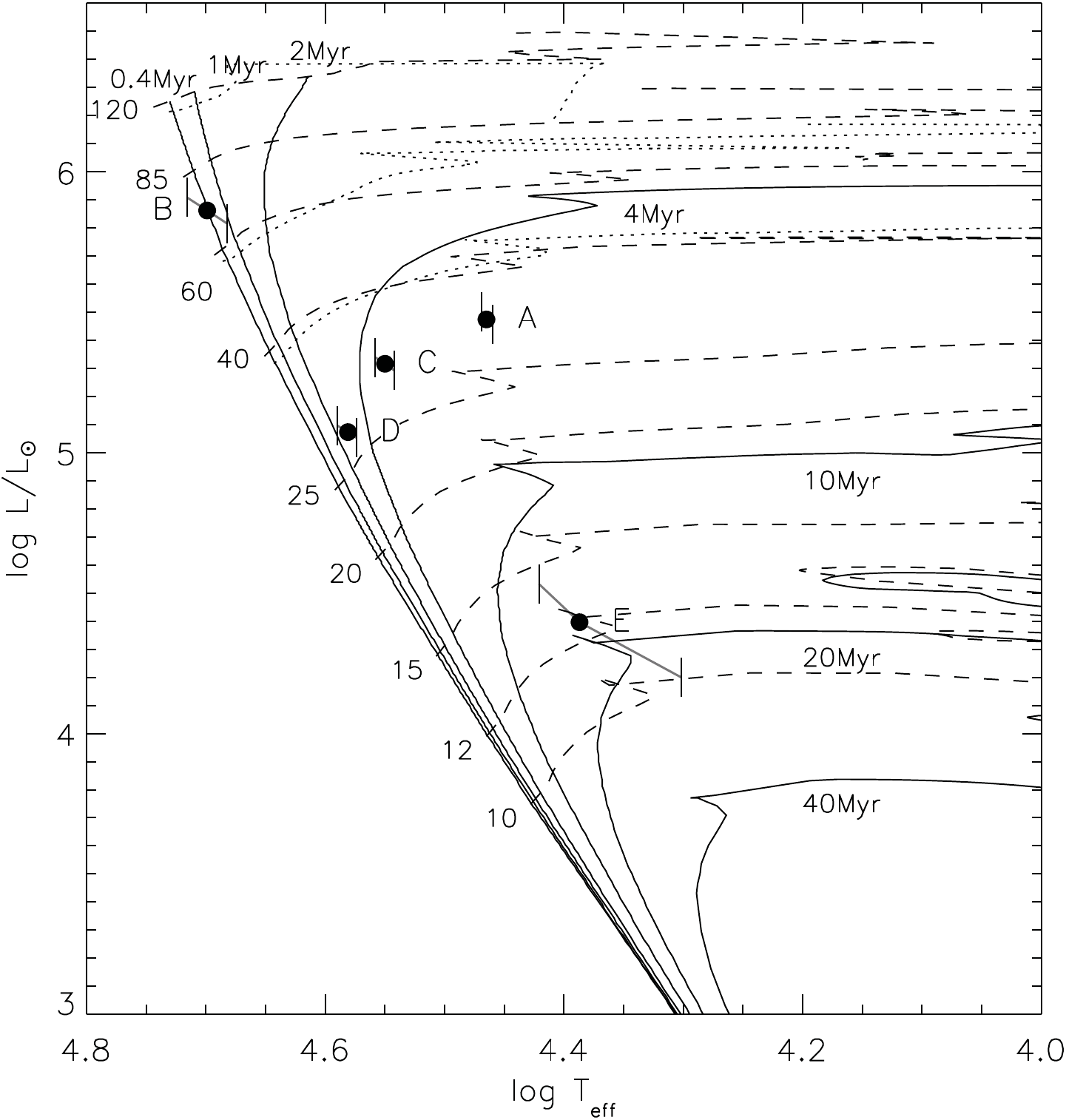}
\caption{H-R diagram of the selected targets, together with evolutionary tracks  for initial masses between 10 and 120~M$_\odot$ (dashed lines) and isochrones for ages between 0.4~Myr and 40~Myr
(solid lines) from the Geneva models of \citet{schaerer+1993} for LMC metallicity. The dotted lines are evolutionary models for stars with rotation, for masses of  40, 60 and 120~M$_\odot$ and LMC metallicity, from \citet{meynet-maeder2005}.
\label{fig:HRD}}
\end{figure*}

We derive the bolometric luminosity from the relation $$\log L/L_\odot=0.4\cdot(M_{\rm bol},_\odot-M_V-BC_V),$$ where $M_{\rm bol},_\odot=4.75$ is the absolute bolometric magnitude of the Sun. The  parameters derived for our five stars in this section are given in columns 1 to 4 of Table~\ref{t:param}. The positions of the stars in the H-R diagram, as well as the errors due to both $T_{\rm eff}$ uncertainties (diagonal error bars) and photometry (vertical error bar hats), are shown in Figure~\ref{fig:HRD}.
In the same plot we overlay a set of Geneva evolutionary models for the LMC metallicity ($Z=0.008$). These are from \citet{schaerer+1993} and are computed assuming no stellar rotation. In general, however, stellar rotation plays an important role in the evolution of massive stars \citep{maeder-meynet2000}, by enhancing mass loss, introducing anisotropy, inducing transport of angular momentum and mixing of chemical elements in radiative zones. While these effects influence the stellar evolution in the H-R diagram \citep{meynet-maeder2000}, they are critical mainly for the evolved stages of massive stellar evolution, and in particular the Wolf-Rayet phase \citep{meynet-maeder2003} and the final pre-supernova stages \citep{hirschi+2004}. In Figure~\ref{fig:HRD} we also plot with dashed lines evolutionary tracks from the grid of models by \citet{schaerer+1993} for selected stellar masses. Unfortunately there are no evolutionary models considering rotation, which are available for the metallicity of the LMC and for the entire high-mass stellar range, except of those computed by \citet{meynet-maeder2005}  for masses of 40, 60 and 120~M$_\odot$, which include an assumed rotation of $v_{ini}=300$~km/s. We show also these tracks in Figure \ref{fig:HRD} with green dotted lines.
For clarity we have not plotted the evolved stages for these masses. From the comparison between the two families of tracks shown in the Figure, it is evident that, at the H-R diagram positions of our five stars, the consideration of rotation or not does not affect significantly the derived stellar parameters.

\subsection{Masses and Ages of the stars}
\label{s:massage}
Considering the above, we utilize the grid of \citet{schaerer+1993}, and assign masses and ages to our stars by interpolation. A particular attention was given to the evaluation of the uncertainties associated to these parameters. As shown in Figure \ref{fig:HRD}, the errors in $\log T_{\rm eff}$ and $\log L$ are correlated and asymmetric. Moreover, the probability for the star to be measured in different positions of the H-R diagram within the error may not be uniform. This is for example the case of source E, the measured $\log T_{\rm eff}$ and $\log L$ of which are located over the $H$-exhaustion locus of the evolutionary models, i.e., the turn-around feature of all the isochrones and tracks in the figure at $\log T\sim4.4$. After the H-exhaustion, the evolution of a star is much faster in the H-R diagram, and thus, the chance of detecting a star at colder temperatures than this point is significantly lower than at hotter temperatures. We take into account this effect to normalize our results by introducing a prior to the estimation of the masses of our targets.

\begin{figure*}
\includegraphics[angle=90,width=\textwidth]{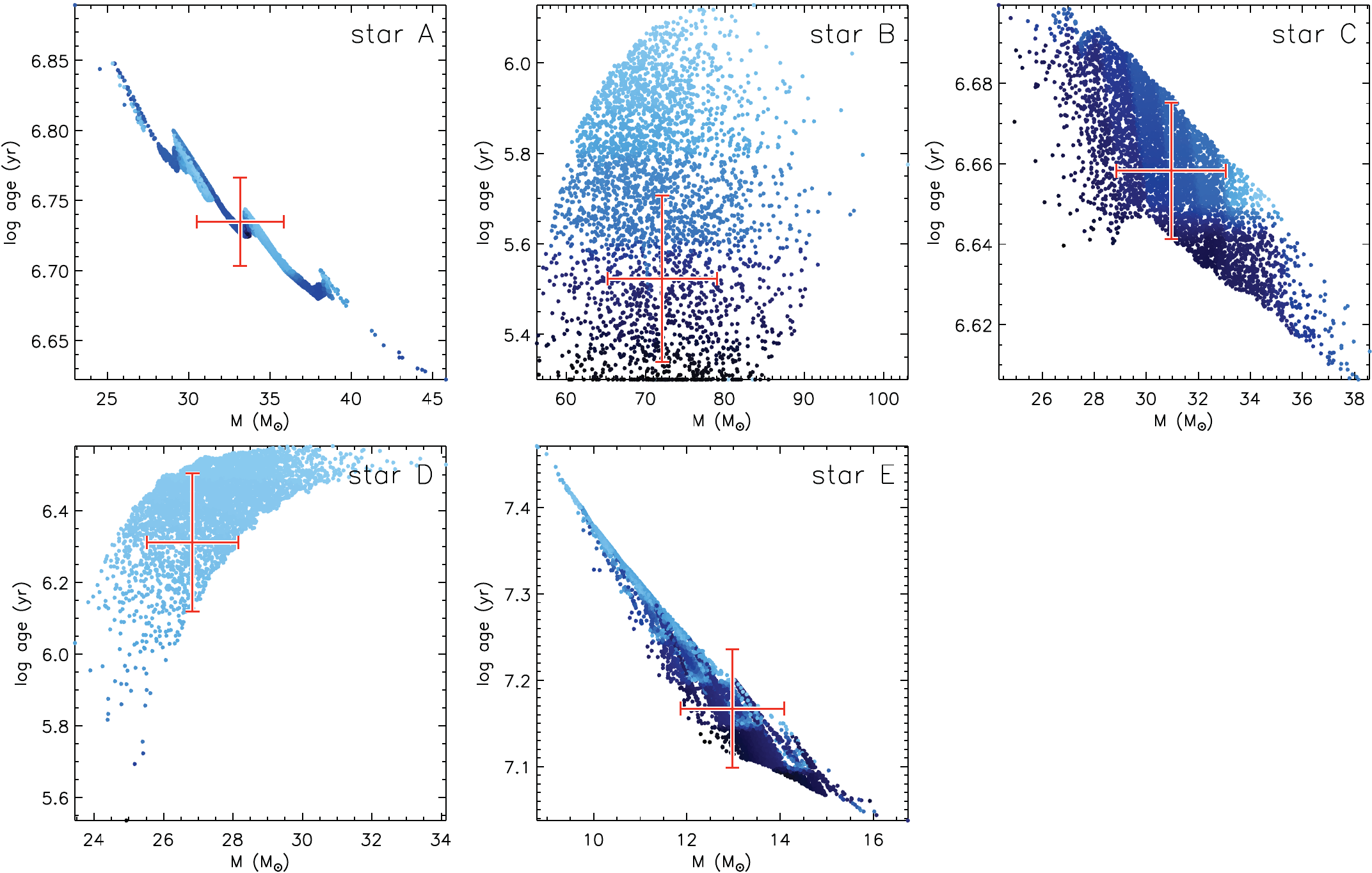}
\caption{Illustration of the MC method to derive the stellar mass and age as well as the associated uncertainties, accounting for the evolution speed in the H-R diagram as a weighting quantity. For every star, the dots represent the transformation in the mass-age plane of the simulated population of stars within the error bars in the H-R diagram. The color is proportional (from dark to light blue) to the rapidity of evolution in the H-R diagram. The crosses represent the derived mean parameters, weighted according to the inverse of the rapidity of evolution.
\label{fig:montecarlo_mass_age}}
\end{figure*}

In order to evaluate the most probable masses and ages for our stars, we proceed with the application of a Monte Carlo (MC) technique, as we describe in the following paragraphs. First, we construct the two-dimensional density distribution of the relative rapidity of stellar evolution on the H-R diagram. For this construction we interpolate the original grid of isochrones and tracks of \citet{schaerer+1993} onto a dense grid with uniform spacing in mass and age. For every point in the new grid, the relative speed of evolution is proportional to the inverse of the distance, on the same track, between neighboring points. This distance depends on the assumed metric, for which we choose a linear scale for both axis of the H-R diagram. Since we are interested in the relative variation of evolution speed between different points in the H-R diagram, the units chosen in both axis to derive the distance between points are arbitrary and equivalent.
Then, for each of our five stars, we generate a well populated number of test stars distributed within the errors shown in Figure \ref{fig:HRD}. We assume a flat distribution in $\log T$, with an equal number of stars on both sides from the central temperature, and a Gaussian distribution in $\log L$, keeping the correlation between the two quantities (the diagonal error bars) as shown in Figure~\ref{fig:HRD}. For each of these simulated stars we assign masses and ages by interpolation on the grid of models, as well as the exact value of relative speed of evolution. Finally, we derive the best mass and age, as well as the uncertainties of these quantities, using weighted statistics, where the weight is the inverse of the evolutionary rapidity. The results are illustrated in Figure \ref{fig:montecarlo_mass_age}, each panel of which represents the mass-age distribution of each of our target stars color-coded according to the weights.

We summarize our results in columns 5 and 6 of Table \ref{t:param}, where we give the most probable masses and ages, and their associated uncertainties for our stars. The influence of weighting the derived stellar parameters on the evolution time scales is evident for stars B and E. For star B, the method tends to favor a slightly younger age than the central value of $\sim0.4$~ Myr shown in Figure \ref{fig:HRD}, while its mass is not significantly affected. For star E, the large uncertainty on the H-R diagram leads to a correlation in the mass-age plane, predicting either a younger more massive, or an older less massive star. Nevertheless, the latter case is less probable than the former, leading to an estimate of the most likely mass of $M=13$~M$_\odot$. This value is larger than that one would obtain by neglecting the errors in the H-R diagram, i.e. $\sim$~12~M$_\odot$ (see Figure \ref{fig:HRD}).

\section{Discussion on the derived ages}\label{section:discussion}

\subsection{Stellar multiplicity}\label{section:binarity}

The stellar parameters derived in the previous section might be biased by the presence of unresolved companions. In fact, most of the early type stars are expected to be formed in binary systems \citep{lada06}, with a binary fraction of the order of $60-100\%$ \citep{zinneckerARAA07}. Given the distance of the Magellanic Clouds\footnote{At the distance of the LMC 1~arcsec corresponds to 50,000~AU}, binary systems generally cannot be resolved.

As mentioned in Section \ref{section:spectral_classification}, for star B we have evidence from the spectrum of an unresolved later-type companion. For the other sources, we are unable to confirm stellar multiplicity based on our spectroscopy. Considering that our spectroscopic observations for stars A, D and E were performed in two epochs, with a time baseline of $\sim2$~months, we attempted to detect evidence of binarity through any radial velocity variations for these stars. We computed the stellar radial velocities, using the cross-correlation technique of \citet{baranne96}. Each stellar spectrum was cross-correlated with a numerical template that was built from the stellar spectrum itself by selecting certain spectral lines. For this propose we rejected blended absorption lines. We used six absorption lines for star A and obtained an accuracy of $\sim$2 km/s. A similar accuracy was also obtained for stars D and E, where we selected four absorption lines.
While we observed no significant radial-velocity variation in stars A and E during the two epochs (November 2008 and January 2009), we detected a linear trend in the radial-velocity variation of star D during this time window. The RV variation between the second (January 2009) and the first epoch (November 2008) is $\sim$9 km/s. This RV variation can be caused by stellar activity or the presence of an unseen stellar companion.

We study now in more detail how unresolved multiple systems can affect our derived masses and ages, and distinguish two cases.

1) The optical luminosity of the companion is significantly smaller than that of the primary. In this case, both the measured spectral type and the optical magnitude are representative of the primary, and thus also the position of the star in the H-R diagram. The presence of the companion, therefore, does not alter the derived parameters of the primary, but affects only the completeness of the stellar sample.

\begin{figure}
\includegraphics[width=\columnwidth]{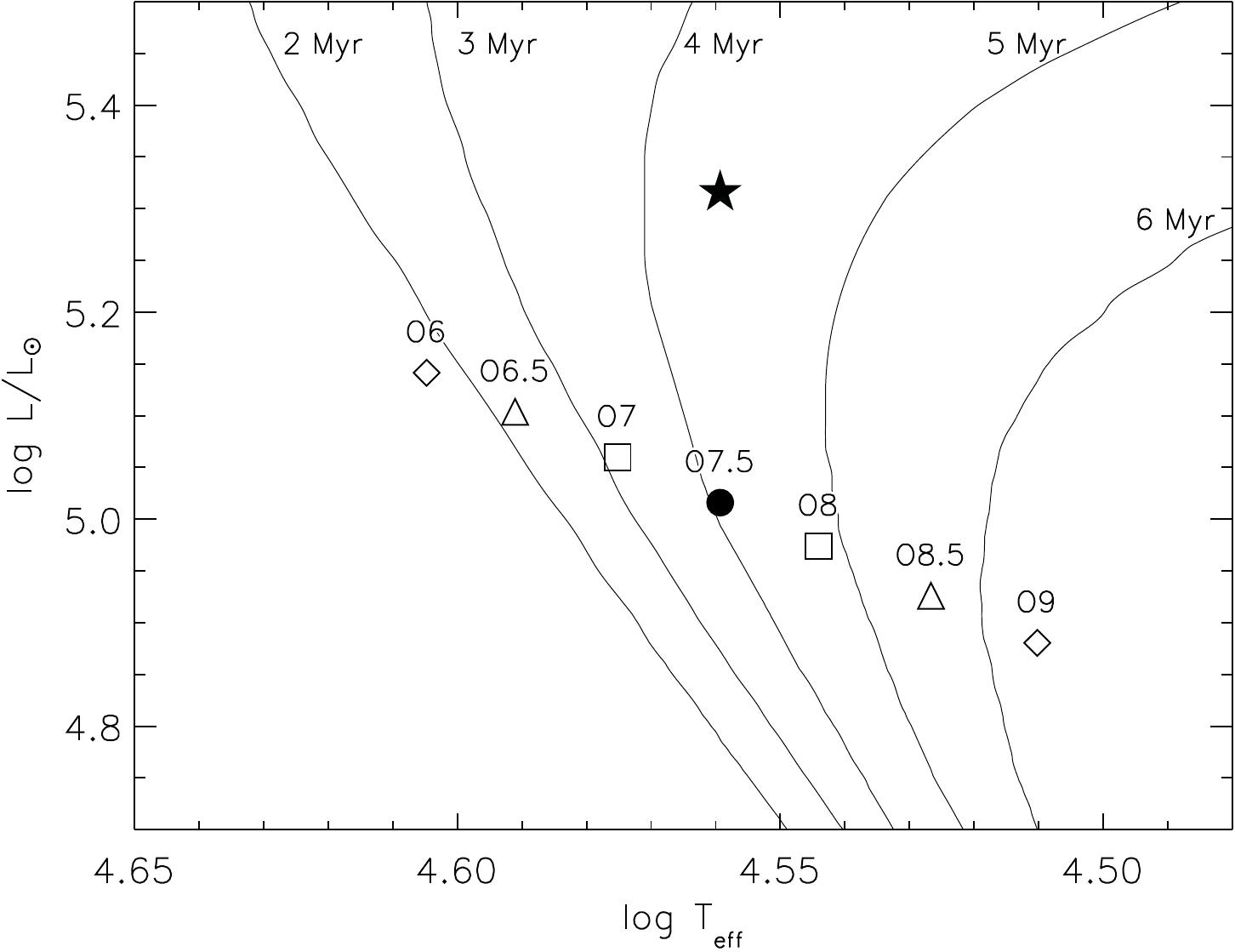}
\caption{Example of how the presence of a case 2\emph{b} unresolved binary system, i.e., two companions of similar optical luminosity but different spectral types,  would bias the derived position of  star C in the H-R diagram. Instead of a single star (denoted by the star symbol), the two companions would be located on symmetric locations (with spectral types O7/O8, O6.5/O7.5, and O6/O9 respectively) at both sides from the average spectral type for this source (denoted by the circle symbol). The diagonal alignment is caused by the increasing BC with temperature. \citet{schaerer+1993} isochrones for ages 2-6~ Myr are overlayed to demonstrate the unrealistic age differences between the companions for a binary system (see text in Section~\ref{section:binarity} for details).  \label{fig:test_starC_binary_offsets}}
\end{figure}

2) The optical luminosity of the companion is comparable to that of the primary. In this case, we further distinguish two cases. \emph{a)} The two stars have identical spectral type: this implies that they are identical companions. They would be located in the H-R diagram at the same $T_{\rm eff}$ we have determined through spectroscopy, and 0.3~dex fainter in $\log L$.
\emph{b)} The two stars have different spectral types. The spectral type we have determined from spectroscopy will be roughly the average of the individual spectral types of the two stars. This scenario, however, would be unrealistic. This is clarified in Figure~\ref{fig:test_starC_binary_offsets}, where we simulate this effect on star C. The two companions with identical optical luminosity but $T_{\rm eff}$ respectively hotter and colder than what we assumed for star C (plotted with a star symbol) would be located in symmetric positions in the H-R diagram (plotted with open symbols) with respect to the case of identical companions (plotted with a filled circle). This would lead to a significant difference in age between the two components, even for a small difference in their spectral types. Given the youth of LH~95, and the fact that its age is comparable to its crossing time (Paper~I), dynamical evolution should not have altered significantly the binarity properties of the systems at the time they were formed. Moreover, since LH~95 is a loose system, crowding should not cause a significant source confusion. Therefore any multiple systems in LH~95 is expected to have be nearly coeval components. As a consequence case 2\emph{b} is not a realistic case for our stars.

We conclude, thus, that the only plausible scenarios are those of cases 1 and 2\emph{a}, i.e., one unseen faint component to our stars, or binaries with companions of nearly identical mass and age, each of them located in the H-R diagram at a luminosity 0.3~dex lower than derived in the previous section. Even in the scenario of case 2\emph{a}, we stress that our derived ages would not be significantly altered, since the $T_{\rm eff}$ of the companions would be very similar to that we have derived from spectroscopy.

\subsubsection{The composite nature of star B}
\label{section:binarity-starB}
The issue of unresolved multiplicity is, on the other hand, more critical for star B, the most massive and also the youngest source in our sample. From our spectroscopy, we found clear evidence of an unresolved colder companion, because of the presence in the spectrum of the \HeI lines. The relative brightness of the secondary component cannot be negligible compared to the primary, because if this would be the case, the ``secondary'' lines would be diluted into the primary continuum,  and  not evident as they are. Since the only major anomaly in the spectrum is the \HeI lines, we argue that the secondary is an early B-type star. By comparing the relative strength of the \HeI lines with respect to the continuum, and considering that the other lines from the primary do not appear significantly dimmed, we estimate that the secondary (either one or more stars) accounts for up to 1/3 of the observed total flux in the B-band wavelength range. Given the very similar intrinsic $(B-V)$ color for all early type stars, a 50\% overestimation of the $B$-band flux results in a 50\% overestimation of the $V$-band flux, and thus of the bolometric luminosity we have derived in Section \ref{section:HRD}. Correcting for this factor (0.17~dex in $\log L$), the mass of this source decreases to $\sim60$~M$_\odot$ and the age reaches the lowest limit covered by the evolutionary models, which predict a ZAMS massive star.

We stress that the secondary component for star B may be a line-of-sight object, not necessarily physically related to the O2 star. Besides the colder secondary we have discussed, we underline that even the main object might be a multiple system consisting of similar components. Given the rarity of such stars, studies of binarity for early O-type stars are limited to a few Galactic objects.
For example, the O2If prototype, HD 93129A is known to harbor 2 very similar stars (HD 93129Aa and HD 93129Ab), resolved thank to HST/FGS observations \citep{nelan2004,nelan2010}. Also Pismis 24-1 (O3.5If + O4III(f)) has been found to consist of at least 3 similar stars, one of the spectroscopic components being an eclipsing binary \citep{maiz-apellaniz2007}. For our star B, however, it would be hard to assume a multiple system, since the individual components would turn out to be even less luminous than the ZAMS (see Figure \ref{fig:HRD}), therefore showing parameters inconsistent with the stellar models.

\subsection{Evidence for an Age Spread}
\label{section:age_spread}
\begin{figure}
\includegraphics[width=\columnwidth]{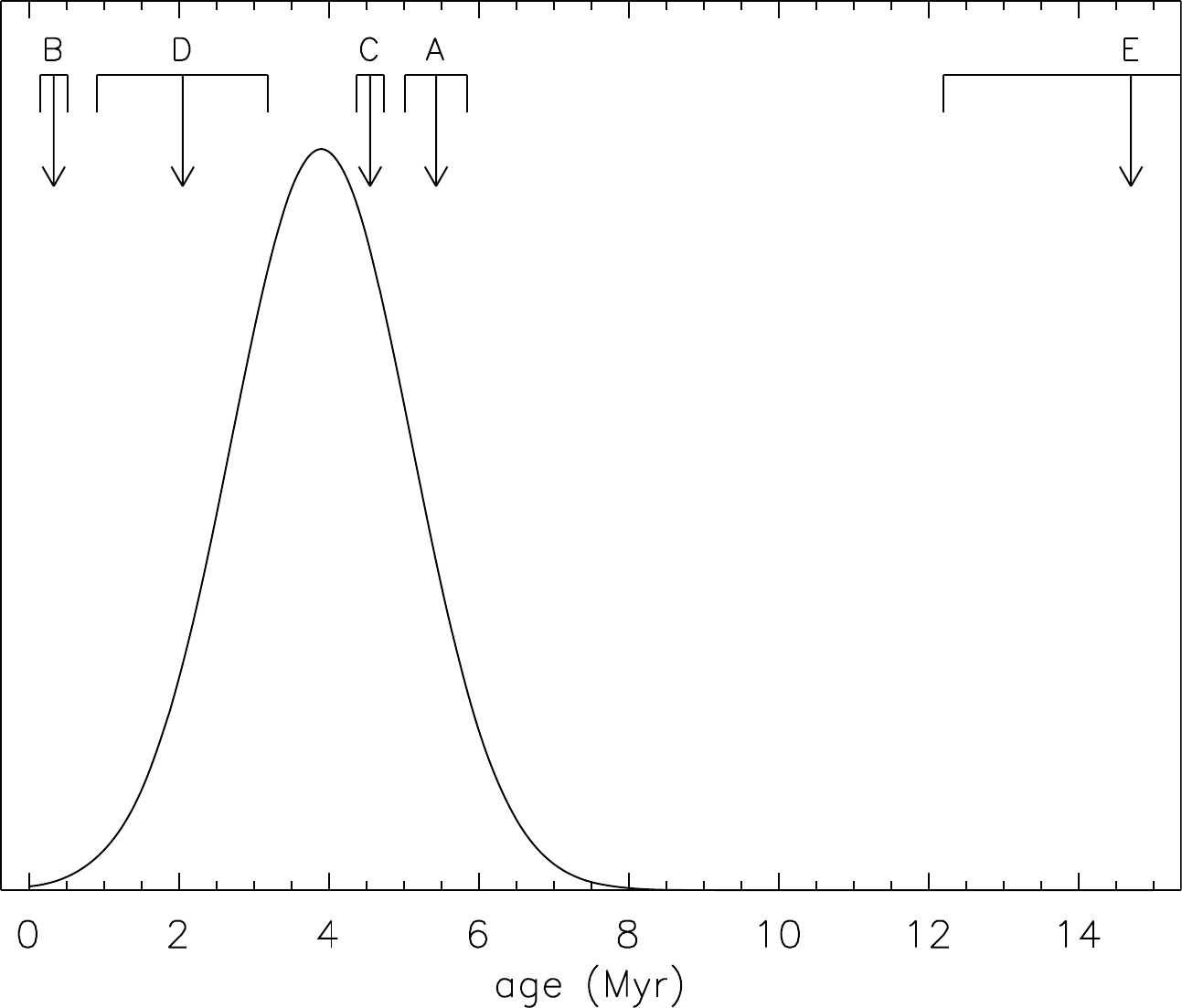}
\caption{The age distribution we derived in Paper~II for LH~95 from the analysis of the low-mass population (gaussian distribution). The arrows, with error bars, indicate the ages and uncertainties of the five massive stars.  \label{fig:agespread}}
\end{figure}

Our findings show that there is a spread in the derived ages for the most massive stars. This result is in agreement with our previous study on the age distribution of low-mass PMS stars in LH~95, based on a statistical analysis of the field subtracted optical CMD (Paper~II). There we determined an average age of the system of about $4$~ Myr, with a confirmed age spread compatible with a gaussian distribution with $\sigma=1.2$~ Myr. In Figure \ref{fig:agespread} we present the derived age distribution for the low-mass population, compared to the derived ages of the five massive members analyzed here. It should be noted that in this plot we preliminarily consider the ages obtained by neglecting any undetected binaries.
From Figure \ref{fig:agespread} it appears that three of our sources  (stars A, C and D) have an age which is compatible with the low-mass age distribution. Actually, the average age for these stars is exactly 4~Myr. On the other hand, star~B appears to be somewhat younger, and star~E is significantly away from the average population.

\begin{figure}
\includegraphics[width=\columnwidth]{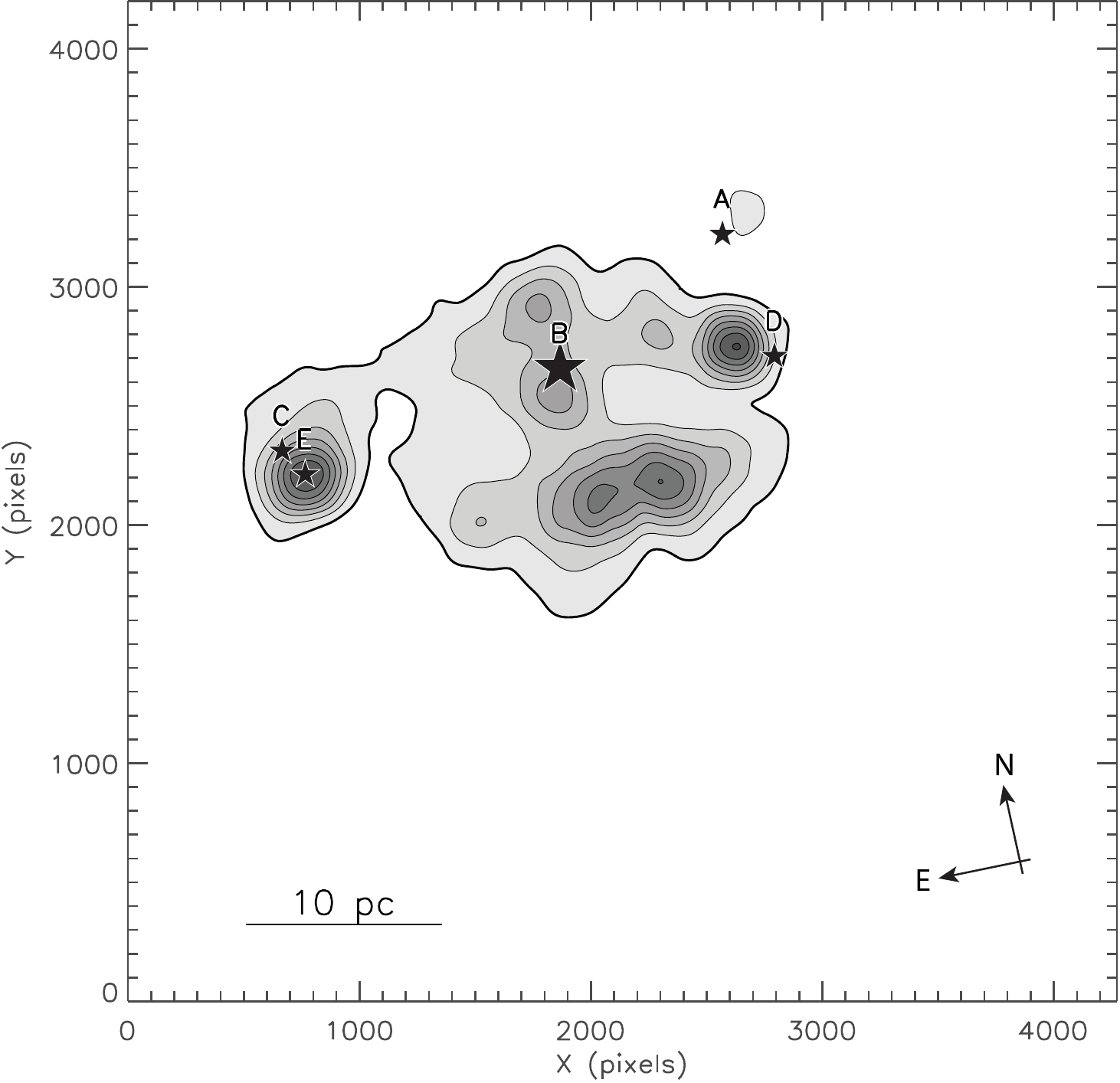}
\caption{Projected positions of our targets, in relation to the spatial density distribution of low-mass PMS stars in the region, analyzed in Paper~I. The thick line delimits the central region of the cluster, where the density of PMS stars is above 3 standard deviations from the sparse background. The contours identify regions of higher concentration of PMS stars. In this study we identified three distinct sub-clusters of such stars (see also Section 3.2 in Paper~I).   \label{fig:density_map_with_targets}}
\end{figure}

Star~E, however, is the faintest source of our sample, and thus has the largest uncertainties; moreover it is located in a particularly crowded sub-cluster, as shown in the stellar surface density map of Figure \ref{fig:density_map_with_targets}. For this source stellar confusion may have contributed significantly in biasing the derived parameters of the star. We assess how the presence of unresolved equal-mass binaries, which would introduce the highest bias in the H-R diagram position of the stars, affects the observed age differences. We find that star E becomes even older, suggesting derived parameters which are even more peculiar. On the other hand, star A would turn out to be 1.4~ Myr older, star D 1.7~ Myr younger, and the age of star C would not be significantly affected by multiplicity. For these three sources, then, the average age remains compatible with that of the low-mass stars even in the case of unresolved binarity, whereas the measured age spread would increase.

Source B, as previously discussed, is the most interesting case. We find that its age is younger than the others, at $3\sigma$ from the average age of the entire cluster, but still covered by the age distribution derived from the low-mass PMS stars. The uncertainty of this age measurement is relatively small. As we mention in Section \ref{section:binarity-starB}, this star appears to be an unresolved composite system, and this implies that the stellar luminosity we have measured (see Table \ref{t:param}) may be overestimated by up to 0.17~dex. This also leads to an even younger age than that we have reported in Section \ref{section:HRD}, whereas the mass of the source
remains quite large ($\sim60-70$~M$_\odot$). In Paper~I we found that LH~95 is a multiple stellar system, consisting of three distinct sub-clusters, identified in terms of the surface density of the low-mass PMS population  in the region (see Figure~\ref{fig:density_map_with_targets}). Moreover, in Paper~II we found that considering the low-mass stars, there is no spatial variation of either average age or age spread throughout the entire region of the cluster, suggesting that the sub-clusters are probably formed coevally. As seen in Figure \ref{fig:density_map_with_targets}, the projected location of source B is very close to the center of the system, but in a region characterized by a relatively low stellar density, and not associated to any of the three main sub-clusters of LH~95. Taking these into account, as well as its young age, it seems that source B  was formed after the rest of the entire stellar population of LH~95, suggesting that the case of this star is rather peculiar.

We can only hypothesize about the origin of a late-born massive star in LH~95. If source B was born in its current position, its formation may have been triggered by the surrounding population of massive stars. However, it is difficult to justify how a massive clump was prevented to collapse in the central part of the system for $\sim 3$~ Myr, when massive star formation was taking place in the surrounding region. Also, the low-mass population in the immediate vicinity of source B does not show evidence of a younger age than the rest of the region.
Another hypothesis is that source B has not formed in its present position. It is known that 10-25\% of the O-type stars are runaways, ejected from their birthplace by dynamical interaction \citep{blaauw1961,poveda67,gies86}. Considering a typical ejection velocity of 40--100~km/s, and given the age of source B, of the order of at most $\sim$~0.5~Myr, it could have formed about 20-50~pc away from the actual position, i.e., still in the vicinity of the rest of the system. From our spectra, however, we do not measure radial velocity variations between star B and the other members. Also, by visual inspection on the NIR {\em Two Micron All Sky Survey} (2MASS) images, or the mid-infrared images obtained with {\em Spitzer Space Telescope}, we do not detect the presence of bow shocks, typically produced by runaway stars. Nevertheless, with no proper motion estimates, we are unable to measure the actual velocity of this source, and therefore confirm it as a runaway star.
We also stress that our lack of knowledge about the geometry of LH~95 along the line of sight limits our ability to isolate the plausible scenarios responsible for our results concerning star B.

\section{The high-mass Stellar IMF} \label{section:IMF}
In Paper~I we derived the IMF of LH~95, well into the sub-solar regime, based on ACS $V$ and $I$-equivalent photometry. We found that the IMF is well approximated by a two-phase power law, with a break point (the ``knee'' of the IMF) at $\sim1$~M$_\odot$. In particular, for stellar masses larger than this point, up to $M \sim$~15~M$_\odot$, we measured an IMF slope $\alpha=3.05$, whereas for stars of sub-solar mass the slope decreases to $\alpha=2.05$. In these units, a \citet{salpeter55} IMF corresponds to a slope $\alpha=2.35$. As mentioned before, the five stars studied here were not included in the analysis of Paper~I, due to saturation in the ACS frames. In this section we complete the stellar IMF of LH~95, by adding the results of our analysis on its high-mass end.

Since the sub-solar part of the IMF will remain unchanged, we limit our investigation to stars found in Paper~I with masses $M>2$~M$_\odot$. In this mass range the stellar sample includes about 110 members. We utilize the method suggested by \citet{imfbias} to derive the mass distribution. These authors find that assuming a constant bin size can lead to misleading results due to the correlation between the number of stars per bin (higher for lower masses) and the assigned weights (from the Poisson statistics). Instead, they demonstrate that assuming a variable bin size, chosen such to have an equal number of sources in each bin, reduces the bias, which remains low also for the limiting case of placing a single star in each bin. Considering the very low number statistics in the high-mass regime in our sample, we choose this counting option for the construction of the IMF.

\begin{figure}
\includegraphics[width=\columnwidth]{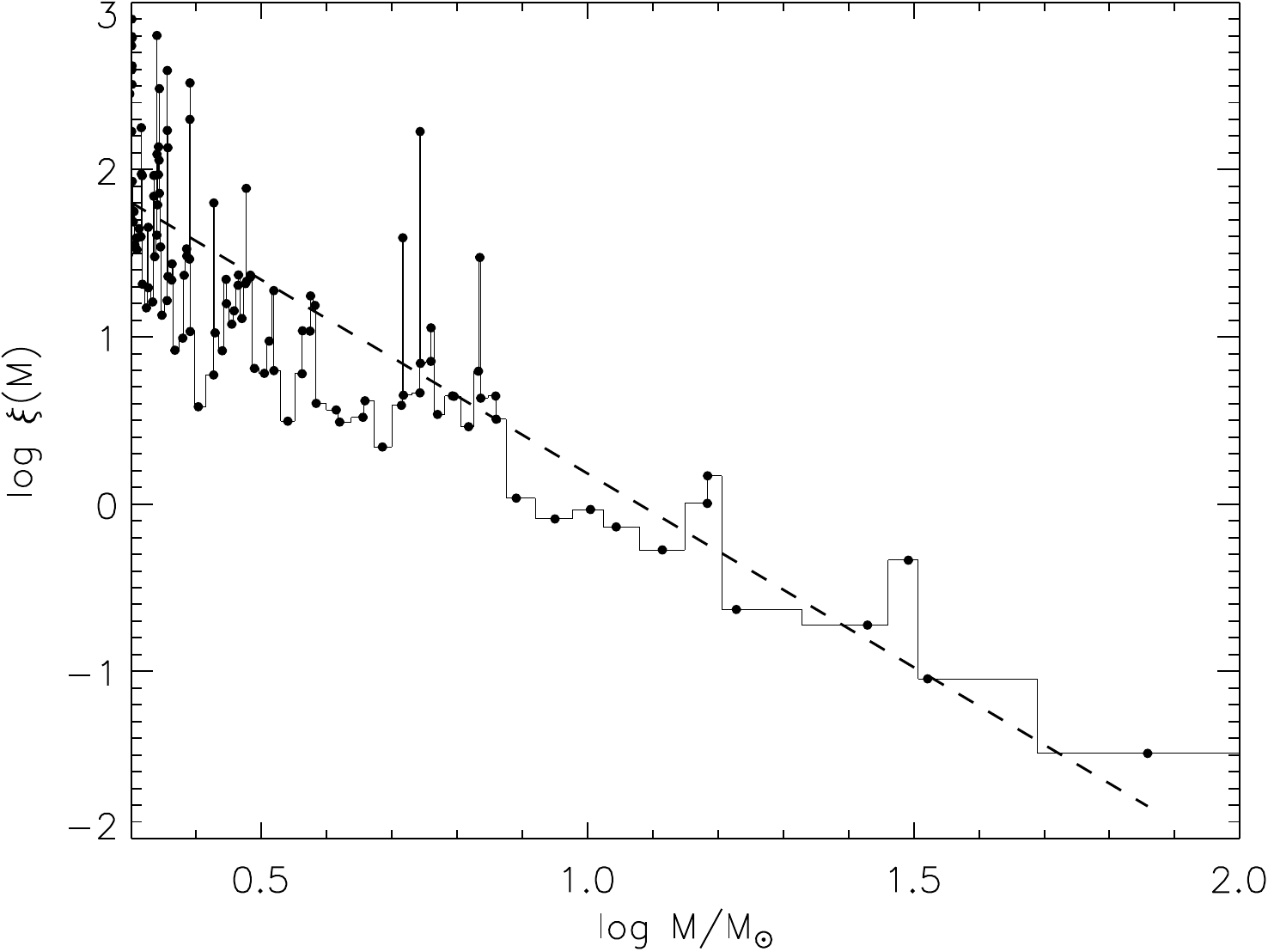}
\caption{IMF for LH~95, limited to the mass interval $M>2$~M$_\odot$. The histogram has a variable bin size, containing one star in each bin. The dashed line represents the best fit of the data. \label{fig:imf}}
\end{figure}

We proceed by first sorting all the stellar members according to their mass. For every star we define the limits of the bin which contains only this star to be the mean (in logarithmic units) between the mass of the star and the preceding and following one respectively. For the most and least massive stars in the sample, the upper and lower boundaries of the bins are chosen to be symmetric from the actual mass of the star. The value of the mass distribution in each bin is then given by the number of stars per bin (1 by definition) divided by the width of the bin. The result is illustrated in Figure \ref{fig:imf}. An additional advantage of this counting method is that, when expressing the IMF in logarithmic units, the weights are identical for each source; thus we can neglect the errors in the fitting process of the IMF, and use a standard linear regression. We find that our IMF is well fitted by a single power low with index $\alpha=2.32\pm0.14$, i.e., identical to a Salpeter IMF. The newly estimated IMF slope turns out to be substantially shallower than that we previously derived without considering the most massive stars of the system, included in our study here.

In Section \ref{section:binarity} we discussed how the presence of unresolved stellar multiplicity may bias the derived ages for the massive members of LH~95. Here, similarly, we investigate if our measured IMF would be altered by binary systems. We consider the limiting case of equal mass binaries, substituting each source with 2 stars of half luminosity, and assign masses by interpolation on the evolutionary models as in Section \ref{s:massage}. For all the other sources with HST/ACS photometry from Paper~I, we assume a binary fraction $f=0.5$, equal mass systems, and assign individual masses assuming the \citet{siess2000} and \citet{girardi02} evolutionary tracks. Finally the IMF is computed and fitted as described above. We find under these assumptions an IMF slope $\alpha=2.0\pm0.1$, therefore slightly shallower than what we obtained neglecting unresolved companions. This result, compatible with similar of such simulations in the literature \citep[e.g.,][]{sagar91}, shows that the consideration of unresolved binarity makes the constructed IMF somewhat shallower, but not enough for its high-mass end to be significantly affected.

\subsection{The relation between the most massive star and the cluster mass}
\label{mmax_mecl}
The correlation between the mass of the most massive star in a cluster, $m_{\rm max}$, and the stellar mass of an entire cluster, $M_{\rm ecl}$, has been long investigated both through empirical studies and analytical modeling \citep[e.g.,][]{larson2003,oey2005,weidner2004,weidner2010}. Its importance is fundamental in several issues, as, for example, it is a critical constrain to the upper mass limit for the formation of a star, and it enables us to investigate whether newborn massive stars in star clusters are randomly drawn from the IMF.

In Paper~I we estimated a total stellar mass for LH~95 of $M_{\rm ecl}\sim2.4\cdot10^3$~M$_\odot$. Accounting for the most massive stars investigated here, we have $M_{\rm ecl}\sim2.5\cdot10^3$~M$_\odot$. Considering also unresolved binaries and a small fraction of diffuse low-mass stars outside the central region, we can constrain $M_{\rm ecl}\sim3.0$--$3.5\cdot10^3$~M$_\odot$. We utilize the model relation $m_{\rm max}$ vs. $M_{\rm ecl}$ from \citet{weidner2004}, derived assuming a maximum stellar mass of 150~M$_\odot$ and the \citet{kroupa2001} IMF. We stress that the assumed stellar upper mass limit does not affect significantly our analysis, given that the LH~95 cluster is not populated enough to host stellar masses near this limit. Moreover, as we discussed in Paper~I, the low mass population of the region follows the Kroupa IMF once the unresolved binaries are accounted for, and the high-mass IMF derived in our study here also agrees with it. The \citet{weidner2004}  model relation predicts a $m_{\rm max}\simeq 70$~M$_\odot$ for $M_{\rm ecl}=3\cdot10^3$~M$_\odot$, and $m_{\rm max}\simeq 80$~M$_\odot$ for $M_{\rm ecl}=3.5\cdot10^3$~M$_\odot$. These values are in very good agreement with our mass estimate of $\sim72$~M$_\odot$ for star B, suggesting that the high-mass population of LH~95 is compatible with a random sampling of stellar masses from its IMF.

On the other hand, our findings are in contrast with the empirical determinations of $m_{\rm max}$ in clusters of different mass. As \citet{weidner2010} point out, carrying out a collective research of cluster masses and maximum stellar mass, there is evidence that the observed $m_{\rm max}$ is statistically lower than that predicted across the entire range of $M_{\rm ecl}$, suggesting the presence of additional mechanisms preventing the formation of the most massive stars in a cluster. These authors show that the relation $m_{\rm max}$ vs. $M_{\rm ecl}$ presents a plateau for clusters with $M_{\rm ecl}$ in the range $10^3$ -- $4\cdot10^3$~M$_\odot$, where the observed $m_{\rm max}\sim25$~M$_\odot$ is well below the predicted value. Therefore LH~95 follows the empirical galactic relation only if star B is not a member of the system. The peculiarly younger age for star B seems to support this idea, and we have already hypothesized that this star could be the product of a subsequent star formation event or even a runaway star (see Section \ref{section:age_spread}). However, both scenarios suggest that star B was born within the extend of the cluster, and therefore there is little argument against the membership of star B to LH~95.

Finally, we mention that \citet{lamb10}, by studying the O-type population of sparse clusters of the SMC, found examples of peculiar $m_{\rm max}$ - $M_{\rm ecl}$ relations. In particular, they reported the presence of massive stars in clusters with a very small $M_{\rm ecl}$, in strong disagreement with both the results of \citet{weidner2010} for the galactic clusters, and the predictions from the IMF. It is clear that our understanding of the formation of massive stars is still somewhat incomplete.

\section{A B-type emission star in LH~95}
\label{section:Be_star}
As discussed in Section \ref{section:spectral_classification}, one star in our sample, star A, shows a prominent Balmer emission. Hydrogen emission lines in B-type stars indicate that they are either PMS stars still forming and emitting through accreting circumstellar disks \citep[classified as Herbig Be stars; see][for a review]{waters98}, or evolved B stars with hydrogen emission, classified as classical Be stars \citep[][]{porter2003}. Classical Be stars are rapidly rotating B-type stars showing a Balmer emission produced in gaseous circumstellar `decretion' disks \citep{porter2003}. Based on its $T_{\rm eff}$ and $L_{bol}$ alone this source could indeed be still in its PMS phase, approaching the MS. However, we exclude this possibility for two reasons. First, the extinction of this star of $A_V=0.88$~mag (Section~\ref{section:HRD}) is very low, implying that it is not embedded, and second its near- and mid-infrared colors, as we present them later, are much bluer than those of typical Herbig Be stars \citep{eiroa2002}.

\begin{figure}
\includegraphics[width=\columnwidth]{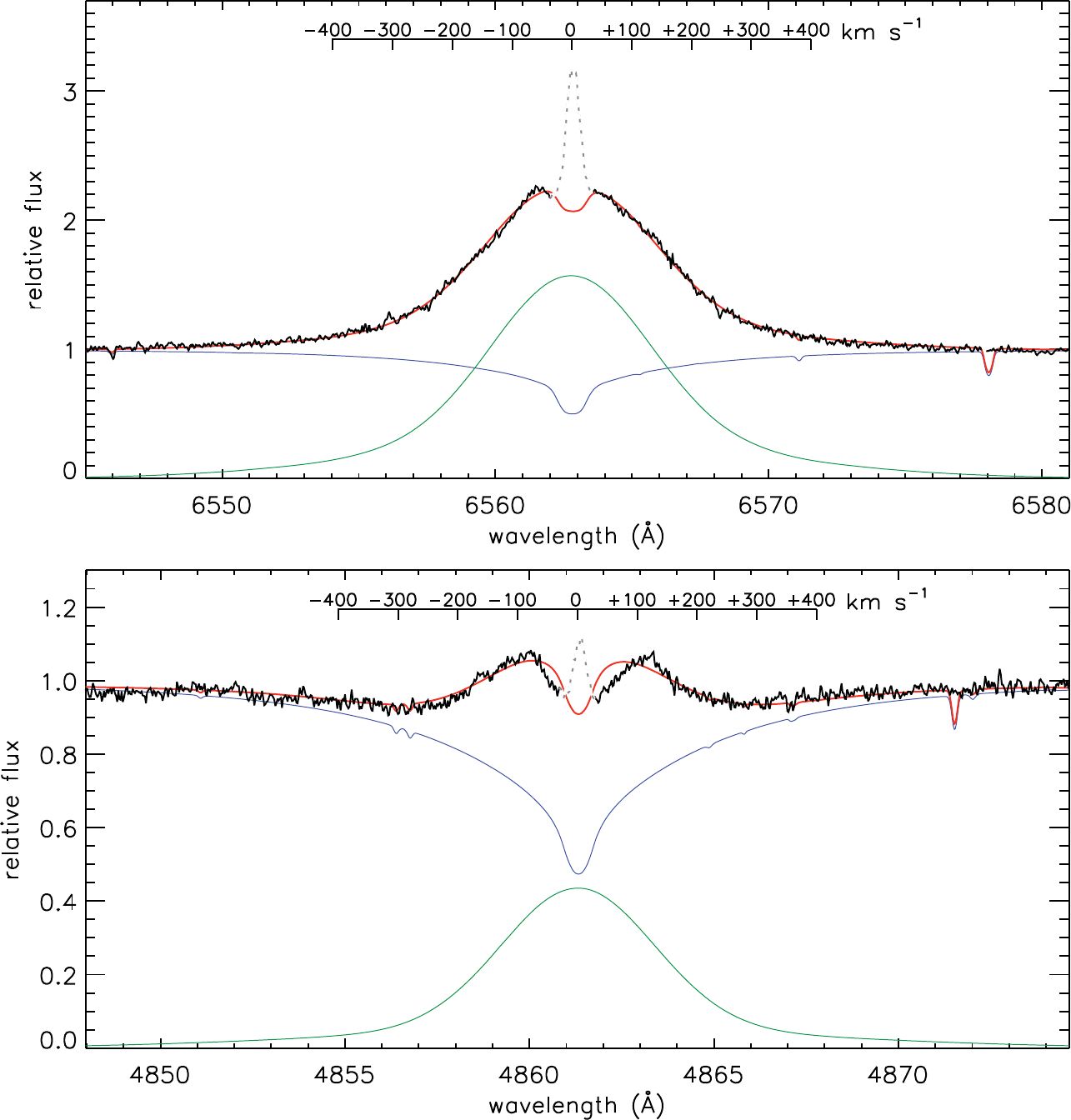}
\caption{H$\alpha$ ({\it top}) and H$\beta$ ({\it bottom}) profiles of the spectrum of star A. The black line represents the observed, continuum normalized spectrum. The narrow emission peaks (grey dashed line) at the center of the lines is the contribution from the diffuse HII region. The blue lines represent the expected photosphere, and the green lines represent the approximated line excess, after the photospheric contribution is being subtracted. The red lines represent the sum of both blue and green lines, demonstrating a good fit to the observed profiles.  \label{fig:starA_balmer}}
\end{figure}

The H$\alpha$ and H$\beta$ profiles from our combined FEROS spectra of star~A are shown in Figure  \ref{fig:starA_balmer}. The H$\alpha$ line presents a very broad, nearly symmetric, emission that reaches relative velocities exceeding 400~Km~s$^{-1}$. At its peak, the net excess is brighter than the local photospheric continuum. The intensity of the H$\beta$ line is  weaker in relation to the continuum. Its width is comparable to that of the H$\alpha$ line, but its profile shows two peaks.  In order to characterize better the excess emission of star~A we subtract the expected photospheric contribution from its spectrum. In Figure \ref{fig:starA_balmer} we show also the approximate shape of the photospheric continuum of the star (blue lines), assuming a TLUSTY \citep{lanz2007} model spectrum with $T_{\rm eff}=29,000$~K, $\log g=4.0$ and $\log Z/Z_\odot=-0.5$. Evidently, these two hydrogen lines are expected to be in absorption from the photosphere. We subtract the photospheric contribution from the observed spectrum and fit the remaining excess with an arbitrary function. Since the emission shows broad wings we find a good match assuming a line profile as the sum of two Gaussian distributions with equal means. The resulting spectrum is also shown for each hydrogen line in Figure \ref{fig:starA_balmer} in green color. While our excess model is possibly oversimplified, we can conclude that the observed double peaked shape of H$\beta$ is not real and can be well reproduced by a broad emission minus a narrower photospheric absorption. Taking the derived emission profiles into account, and given the absence of additional absorption from the circumstellar environment, we classify star~A as a classical ``single line'' Be star, rather than a Shell Be star, i.e., a Be star whose disk is seen edge-on \citep{rivinius2006}.

\begin{figure}
\includegraphics[width=\columnwidth]{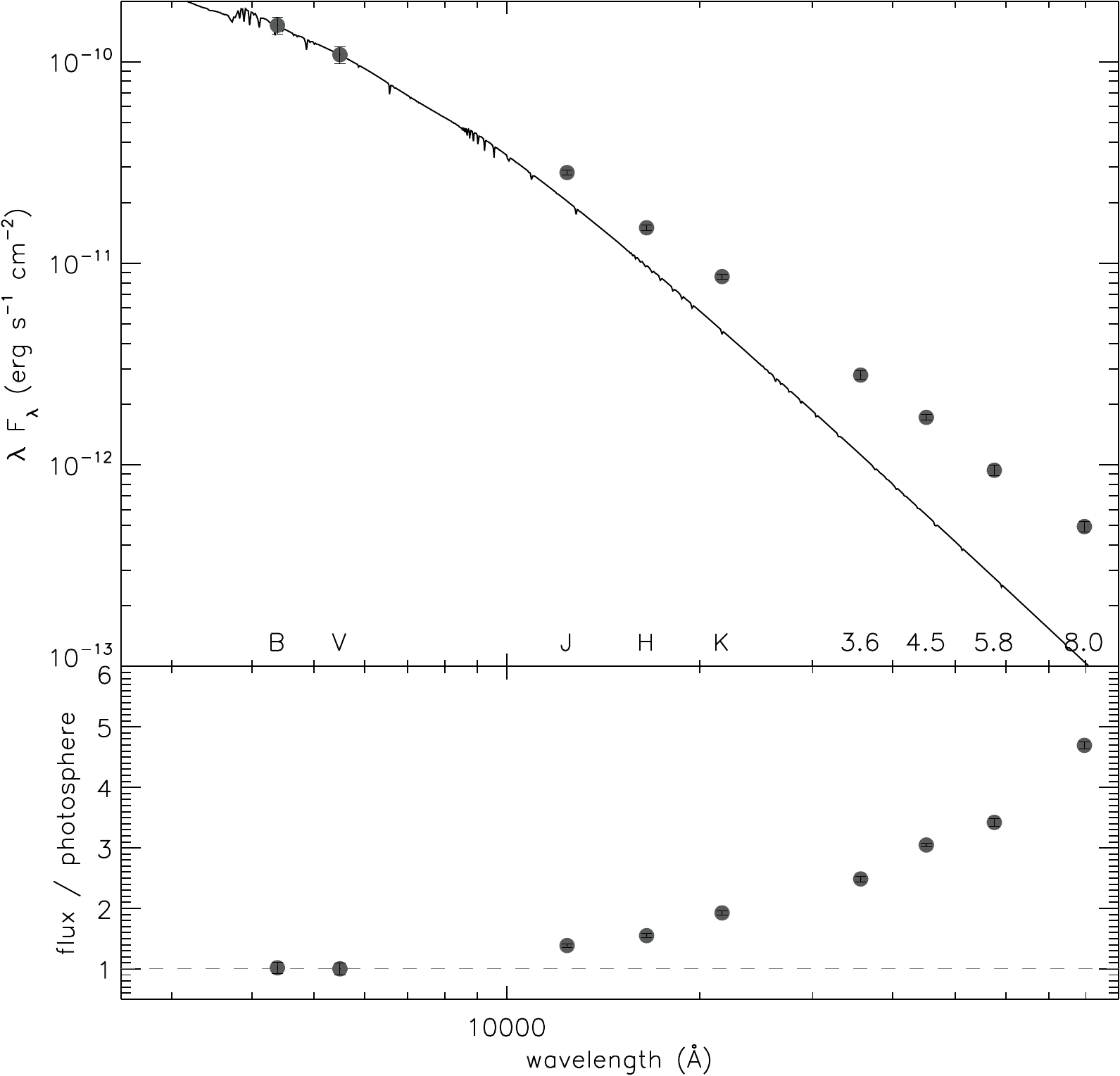}
\caption{{\em Upper panel:} the measured SED of star A based on our optical photometry, 2MASS, and Spitzer/IRAC measurements (dots). The black lines represents the expected photosphere ($T_{\rm eff}=29,000$~K) to which we applied the measured extinction of $A_V=0.88$~mag. {\em Lower panel:} The relative ratio in the fluxes (observed divided by photosphere). \label{fig:starA_SED}}
\end{figure}

Fast stellar rotation is a typical property of Be stars and it is a significant contributor to the generation of the circumstellar medium surrounding these objects. The typical rotation rates of Be stars may be of several hundreds of km~s$^{-1}$, a significant fraction of the critical rate. Our spectrum of star A, however, does not show significant line broadening due to stellar rotation. We have quantified the (projected) rotational velocity by considering the synthetic TLUSTY spectrum assumed before, convolved with line spread functions of different values of $v \sin i$. We have then compared the width of the \HeI with the observed ones. Assuming for simplicity rigid rotation, we obtained $v \sin i \sim 60$~km~s$^{-1}$. This is a small value for this class of stars, and suggests that the stars' rotation axis is not very inclined from the line of sight. This is compatible with our previous argument that the circumstellar disk is not seen edge-on given the lack of Balmer self-absorption.

Classical Be stars are also known to show an infrared excess, commonly interpreted as free-free and free-bound emission of the disk \citep{gehrz74,porter2003,Touhami2011}. In order to verify this characteristic for star~A, we collected near- and mid-infrared photometry for this source. Specifically, we retrieved from the NASA/IPAC Infrared Science Archive\footnote{IRSA is available at http://irsa.ipac.caltech.edu/} (IRSA)  $JHK$ near-IR photometry for star~A from the {\em Two Micron All Sky Survey} (2MASS) and mid-IR photometry in 3.6, 4.5, 5.8 and 8.0~$\mu$m from the {\em  Infrared Array Camera} \citep[IRAC;][]{fazio04} on-board the {\em Spitzer Space Telescope} \citep{werner04}. The resulting spectral energy distribution (SED) is shown in Figure \ref{fig:starA_SED}, together with the expected photosphere after adding an extinction of $A_V=0.88$~mag, as estimated in Section~\ref{section:HRD}.  The comparison between the SED and the model photosphere confirms the presence of infrared excess for this source. We stress that the near- and mid-IR part of this SED could not be reproduced by any B-type photospheric model even if we would assume a significantly different stellar $T_{\rm eff}$ than that of star~A. The reason lies on the fact that at these wavelengths the spectrum is in the Rayleigh-Jeans regime and so the SED slope does not depend on $T_{\rm eff}$. Moreover, given the negligible effect of dust extinction at these wavelengths -- and the fact that we do measure a very small $A_V$ -- an inaccurate reddening estimation can neither justify the observed SED. As a consequence, the observed IR excess of star~A is real.

Be stars may also present line variability because of changes in their disk state \citep{mcswain2009}. Since our optical spectroscopy for this source was obtained in two epochs (see Table \ref{t:obs}), we searched for variations in the Balmer emission by inspecting the individual exposures from each epoch. Within the $\sim 2$ month baseline between the observations, we do not detect any change in the shapes, widths, or intensity of neither the H$\alpha$ nor the H$\beta$ emission, as they are shown in Figure \ref{fig:starA_balmer}. Finally, it is interesting to note that the inferred age of $\sim4$~ Myr for this star (see Table \ref{t:param} and Figure \ref{fig:HRD}) relies on the assumption that this source is a classical Be star, evolving off the main sequence, and not a forming PMS star.


\section{Conclusions}
\label{section:conclusions}
This paper complements our investigation of the stellar OB association LH~95 in the LMC. In our previous studies (Papers~I and II), based on deep HST/ACS imaging, we investigated the intermediate- and low-mass PMS population of the system down to $0.4~ M_\odot$. We derived the corresponding IMF, which is found to be compatible with the \citet{kroupa2001} IMF (Paper~I), we determined the average cluster age to be $\sim4~$ Myr and confirmed an age spread of the order of 2-4~ Myr (see also Paper~II). We did not find evidence of spatial variations neither in the IMF nor the age distribution.

In the present study we focus on the brightest sources in the region, which were saturated in our previous HST imaging.  We use ground-based high resolution optical spectroscopy complemented by previous optical photometry to derive the stellar parameters of the five most massive stars, named after the letters A to E, completing thus our census of the LH~95 stellar population across the entire mass spectrum. We perform spectral classification of the sources in terms of the identification of specific spectral lines.  We assign masses and ages to these stars, and we discuss the related uncertainties, based on comparison of their H-R diagram positions with evolutionary models. We identify an age distribution  among the stars, and we use their mass measurements to complete the stellar IMF of LH~95 toward the high-mass regime.
We summarize our findings to the following:

\begin{itemize}

  \item[--]  The most massive star in the region, star~B, is an O2 giant with a mass of $\sim60-70$~M$_\odot$. The case of this star is very interesting. Specifically, it appears to be  somewhat younger  ($<1$~ Myr) than the rest of the system. In addition, assuming that the established relation between the maximum stellar mass in a cluster and the total mass of the cluster represents the vast majority of clusters, star~B is  more massive  than expected, given the total mass of the stellar system ($\sim 3\cdot10^3$~M$_\odot$). As such, LH~95 appears to be an outlier in this relation. Based on our analysis, we exclude any major bias that might have lead us to derive inaccurate stellar parameters for this source.

  \item[--]  The brightest star in the sample, star~A, is the second most massive. It is a $\sim 33$~M$_\odot$ B0.2 giant with significant hydrogen emission. Specifically, it shows very broad H$\alpha$ and H$\beta$ lines in emission, originating from a gaseous circumstellar disk, as well as clear near- and mid-infrared excess. Considering its low extinction and blue infrared colors, we assess that star~A is not a Herbig Be PMS star evolving toward the MS, but a classical Be star, evolving away from it.

  \item[--]  The average ages and age spread of the most massive stars are consistent with those previously derived by us for the low-mass PMS populations. Exception to this behavior is star~E, which appears  significantly older than the rest of the stars. We assign this peculiarity to the severe crowding by low-mass stars in its immediate surroundings that affects its photometric measurements, and its low brightness, which introduces large uncertainties in our measurements.

 \item[--] The complete stellar IMF of LH~95, from the most massive stars down to the intermediate mass regime with \gsim~2~M$_\odot$ follows a Salpeter slope. Together with the low-mass stars, therefore, this IMF is compatible with a Kroupa IMF.

   \end{itemize}

\section*{Acknowledgments}
We thank Nolan Walborn for his very helpful feedback on the spectral classification.
The authors acknowledge the Max-Planck Society
(MPG) and the Max-Planck Institute for Astronomy
(Heidelberg, Germany) for the telescope time.
D.A.G. kindly acknowledges financial support by the German Aerospace
Center (DLR) and the German Federal Ministry for Economics and
Technology (BMWi)  through grant 50~OR~0908, and by
the German Research Foundation (DFG) through grant GO~1659/3-1.


\end{document}